\documentclass[twocolumn,showpacs,preprintnumbers,amsmath,amssymb,aps,prl]{revtex4}
\usepackage{graphicx}
\usepackage{textcomp}

\usepackage{amsfonts}
\usepackage{amssymb}
\usepackage{amsmath}
\usepackage{calc,graphicx,color,bm}

\def\be{ \begin{equation} }
\def\ee{ \end{equation} }
\def\bea{ \begin{eqnarray} }
\def\eea{ \end{eqnarray} }
\def\bse{ \begin{subequations} }
\def\ese{ \end{subequations} }

\def\e{e}

\def\to{\rightarrow}

\def\half{\tfrac12}

\begin{document}

\author{Boyan T. Torosov}
\altaffiliation{Permanent address: Institute of Solid State Physics, Bulgarian Academy of Sciences, 72 Tsarigradsko chauss\'{e}e, 1784 Sofia, Bulgaria}
\affiliation{Dipartimento di Fisica, Politecnico di Milano and Istituto di Fotonica e Nanotecnologie del Consiglio Nazionale delle Ricerche, Piazza L. da Vinci 32, I-20133 Milano, Italy}
\author{Giuseppe Della Valle}
\affiliation{Dipartimento di Fisica, Politecnico di Milano and Istituto di Fotonica e Nanotecnologie del Consiglio Nazionale delle Ricerche, Piazza L. da Vinci 32, I-20133 Milano, Italy}
\author{Stefano Longhi}
\affiliation{Dipartimento di Fisica, Politecnico di Milano and Istituto di Fotonica e Nanotecnologie del Consiglio Nazionale delle Ricerche, Piazza L. da Vinci 32, I-20133 Milano, Italy}
\title{The imaginary Kapitza pendulum}
\date{\today }

\begin{abstract}
We extend the theory of Kapitza stabilization within the complex domain, i.e. for the case of an imaginary oscillating potential. At a high oscillation frequency, the quasi-energy spectrum is found to be entirely real-valued, however a substantial difference with respect to a real potential emerges, that is the formation of a truly bound state instead of a resonance. The predictions of the Kapitza averaging method and the transition from a complex to an entirely real-valued quasi-energy spectrum  at high frequencies are confirmed by numerical simulations of the Schr\"{o}dinger equation for an oscillating Gaussian potential. An application and a physical implementation of the imaginary Kapitza pendulum to the stability of optical resonators with variable reflectivity is discussed. 
\end{abstract}

\maketitle


\section{Introduction}
The Kapitza (or dynamical) stabilization effect refers to the possibility for a classical or quantum particle to be trapped by a
rapidly-oscillating potential in cases where the static potential cannot \cite{Landau}. It was first explained and demonstrated in classical physics by Pyotr Kapitza in 1951, who showed that an inverted pendulum can be stabilized by the addition of a vertical vibration \cite{Kapitza}.  Later nonlinear and quantum analogues of this phenomenon were studied in several papers \cite{Nonlinear Kapitza 1, Nonlinear Kapitza 2, Quantum Kapitza 1,Quantum Kapitza 2} and found important applications, for example in Paul traps for charged particles  \cite{PTrap}, in driven bosonic Josephson junctions \cite{bos}, and in nonlinear dispersion management and diffraction control of light in optics \cite{Nonlinear Kapitza 1, Ciattoni}.  The main result underlying Kapitza stabilization is that the motion of a classical or quantum particle in an external rapidly oscillating potential can be described at leading order by an effective time-independent potential, which shows a local minimum (a well) while the non-oscillating potential did not. While in the classical description the oscillation-induced  potential well introduces a locally-stable fixed point of the motion, in the quantum description stabilization is imperfect since the effective potential does not sustain truly bound states, rather resonance states with a finite lifetime owing to quantum tunneling \cite{Quantum Kapitza 2}. 

Kapitza stabilization  has been studied so far only when the potential is real. In recent years the physics and applications of non-Hermitian systems has received an increasing attention \cite{NH1,NH2}, especially in the context of $\mathcal{PT}$-symmetric systems \cite{PT}. Such systems possess novel and unexpected physical features. For instance they can be used for faster-than-Hermitian evolution in a two-state quantum system \cite{Faster}. Complex potentials can be experimentally realized in different systems, including open two-level atomic systems interacting with near resonant light \cite{MW} and optical structures with gain and loss regions \cite{OP}.

In this work we extend  the problem of the Kapitza pendulum to non-Hermitian Hamiltonians, i.e. to the case of an imaginary rapidly-oscillating potential. In the classical description, an ``imaginary'' oscillating inertial force introduces two (rather than one) stable fixed points. In the quantum description,  the quasi-energy spectrum of the Hamiltonian turns out to be entirely real-valued in the high-frequency regime. As compared to a real potential, the main distinctive feature here is that the imaginary oscillating potential sustains a truly (Floquet) bound state instead of a resonance state, i.e. stabilization is perfect for the imaginary oscillating potential while it is imperfect for the real oscillating potential. 

\begin{figure*}[tb]
\centering \includegraphics[width=2\columnwidth]{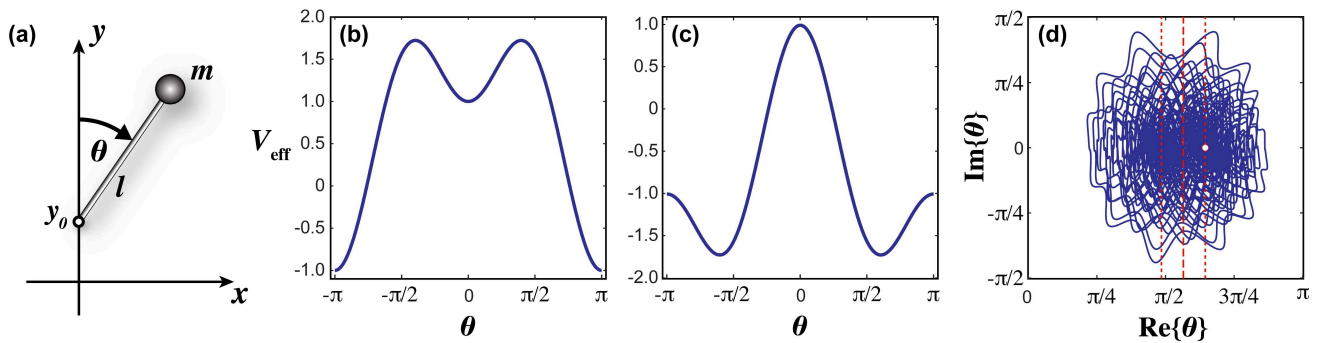}
\caption{(Color online) Complex extension of the Kapitza pendulum. (a) A rigid pendulum with a vibrating pivot $y_0(t)=A \cos (\omega t)$. (b) Effective potential $V_{\rm eff}$ (in units $mgl$) for a real inertial force ($A^2\omega^2/gl=10$). (c) Effective potential for an ``imaginary''  inertial force ($A^2\omega^2/gl=-10$). (d) Complex trajectory for $A/l=0.1\i$ and $l\omega^2/g=1000$. The initial condition is $\theta(0)=\arccos{2gl/A^2\omega^2} +0.25\approx 2.02$, $\dot \theta(0)=0$, and is marked by the red point. The dashed line shows the equilibrium position and the two dotted lines show the turning points as determined by the effective potential shown in (c).
}\label{portrait}
\end{figure*}

\section {Imaginary Kapitza pendulum} We start with a short overview of the classical Kapitza-pendulum problem \cite{Landau}. Let us consider a rigid pendulum of mass $m$ and length $l$, and let us assume that its pivot point is vibrating rapidly in vertical direction, $y_0 = A \cos (\omega t)$ [see Fig.~\ref{portrait}(a)]. The Lagrangian of the system in the inertial reference frame is given by $\mathcal{L}=E_{\text{k}}-V_{\text{p}}$, where
\bse
\begin{align}
&E_{\text{k}}=\frac{ m l^2\dot{\theta}^2}{2}+m A l\omega\dot{\theta}\sin\omega t\sin\theta+\frac{ m A^2\omega^2}{2} \sin^2 (\omega t) ,\\
&V_{\text{p}}=m g(l\cos\theta+A\cos\omega t) ,
\end{align}
\ese
are the kinetic and potential energies, respectively, $g$ is the gravitational acceleration and $\theta$ is the angle between the pendulum and the $y$-axis. The equation of motion for the angle $\theta$, as obtained from the Euler-Lagrange equations, reads
\be
\ddot{\theta}=\frac{\sin\theta}{l}(g-A\omega^2\cos\omega t).
\ee
Following the idea of Kapitza \cite{Kapitza}, the variable $\theta$ can be written as the sum of fast and slowly-varying variables, namely  $\theta(t)=\theta_0(t)+\xi(t)$, where $\theta_0$ is a slowly-varying function over one oscillation cycle and $\xi$ is the rapidly oscillating part. The latter is given by $\xi=(A/l)\sin\theta\cos\omega t$. After writing the equation of motion for the ``slow'' component $\theta_0$ and averaging over the rapid oscillations, we can derive an effective potential $V_{\text{eff}}$ for $\theta_0(t)$ \cite{Landau},
\be\label{VeffClassical}
V_{\text{eff}}= m g l \left(\cos\theta + \frac{ A^2 \omega^2}{4 g l} \sin^2\theta \right) .
\ee
We see that, for a proper choice of the parameter values, namely for $A^2 \omega^2>2gl$,  $V_{\text{eff}}$ has two minima, a local minimum at $\theta=0$ and a global minimum at $\theta =\pm\pi$; see Fig.~\ref{portrait}(b). This result shows that  the upper vertical stationary position $\theta=0$, which is an unstable point  for an ordinary pendulum, can be stabilized when the suspension point  of the pendulum vibrates rapidly.  We can generalize the Kapitza idea into the complex domain by allowing the underlying potential to become complex-valued.  
Complex extensions of classical mechanics  and the strange dynamics of a classical particle subject to complex forces and  moving about in the complex plane have been studied in several works \cite{Complex trajectories}, especially in the context of $\mathcal{PT}$-symmetric classical mechanics theory \cite{PT}. In our case, we extend the Kapitza pendulum into the complex domain by allowing the non-inertial force $m \ddot{y}_0$ to become imaginary.  This corresponds to a change of the amplitude $A$ of oscillation into $i A$, which leads to a flip in the sign of the second term in Eq.~\eqref{VeffClassical} and to changing the effective potential $V_{\rm eff}$. Noticeably, the effective potential  remains real-valued in spite of the imaginary non-inertial force. As compared to the real Kapitza pendulum, in the imaginary Kapitza pendulum the effective potential has two global minima in the nonintuitive positions $\theta=\pm\arccos{2gl/A^2\omega^2}$, see Fig.~\ref{portrait}(c). Even though the effective potential for the slow variable $\theta_0$ is real-valued, the trajectory of $\theta(t)$ occurs in the complex plane owing to the rapidly-varying component $\xi(t)$,  which is imaginary. An example of such a complex trajectory for the imaginary Kapitza pendulum is shown on Fig.~\ref{portrait}(d). The figure clearly shows the oscillatory motion in the complex plane around one of the fixed stable points of $V_{\rm eff}$.

Let us now consider the imaginary Kapitza pendulum problem in the quantum mechanical framework. We consider the motion of a quantum particle in a one-dimensional time-dependent potential $V(x,t)$, which is described by the Schr\"{o}dinger equation (with $\hbar=m=1$)
\be\label{Schr}
i \partial_t \psi(x,t) = - \half\partial_{x x} \psi(x,t) + V(x,t) \psi(x,t) \equiv \hat{H}(x,t) \psi(x,t) .
\ee
As in Ref.~\cite{Quantum Kapitza 2}, the external potential $V(x,t)$ is taken of the form
\be\label{eq5}
V(x,t)=W(x)\cos\omega t ,
\ee
where $W(x)$ is assumed to vanish at $x\to \pm\infty$ and $\omega$ is the modulation frequency. 

\begin{figure}[b]
\centering \includegraphics[width=0.9\columnwidth]{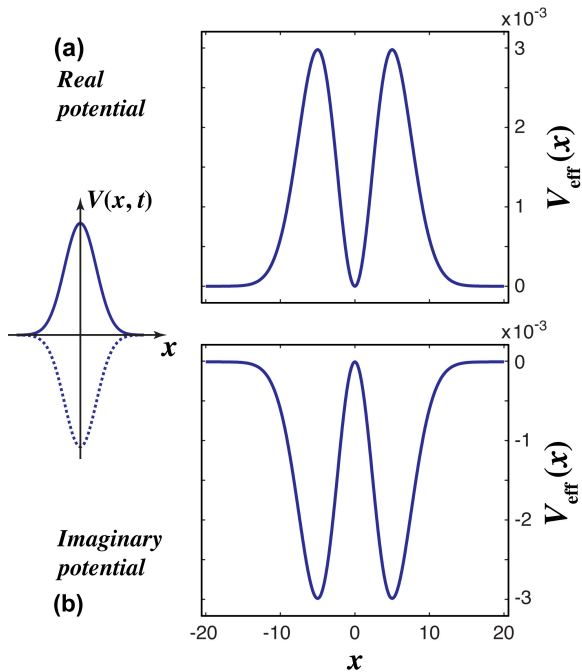}
\caption{(Color online) External time-dependent potential $V(x,t)$ with a Gaussian shape (left) and effective time-independent potential for (a) a real potential, and (b) a purely imaginary potential (right). Note that in the latter case the effective potential is a double well. Parameter values are $V_0=9$, $\beta=0.02$, and $\omega=10$.
}\label{potential}
\end{figure}

Similarly to the classical case, in the high modulation regime such a potential can be approximated by an effective time-independent potential \cite{Quantum Kapitza 1, Quantum Kapitza 2}
\be\label{Vapprox}
V_{\text{eff}}(x) \approx \frac{1}{2\omega^2} \left\langle\left( \frac{\partial V(x,t)}{\partial x} \right)^2\right\rangle  = \frac{1}{4\omega^2} \left( \frac{\partial W(x)}{\partial x} \right)^2 ,
\ee
where the brackets $<>$ denote time average and the error of this approximation is $\mathcal{O}(\omega^{-4})$.
If $W(x)$ is a real function, then $V_{\text{eff}}(x)>0$, while if $W(x)$ is an imaginary function, then $V_{\text{eff}}(x)<0$, with $V_{\rm eff}(x) \rightarrow 0$ as $ x \rightarrow \pm \infty$ in both cases.  In the first case, because of the quantum tunneling, bound states are unlikely and the energy spectrum of the Hamiltonian is continuous \cite{note}; only resonance states may exist, as shown in Ref.~\cite{Quantum Kapitza 2}. In the second case, however, it is possible to obtain truly bound states.
As an example, let us consider a Gaussian potential $W(x)$, 
\be\label{V}
W(x) = i V_0 \exp(-\beta x^2),
\ee
where $V_0$ and $\beta$ are assumed real. According to Eq.~\eqref{Vapprox}, such potential is approximated by the effective time-independent potential
\be\label{Veff}
V_{\text{eff}}(x) \approx  -\frac{V_0^2\beta^2}{\omega^2}x^2\e^{-2\beta x^2},
\ee
 We see that, despite the fact that $V(x,t)$ is purely imaginary zero-average potential, the frequency modulation leads to an effective real potential. In Fig.~\ref{potential} we show schematically the correspondence between a frequency-modulated potential and the effective time-independent potential, both in the case of purely real [panel (a)] and purely imaginary [panel (b)] potentials. We see that in the case of a real potential, the corresponding effective potential is a double barrier, while in the case of an imaginary original potential, the effective potential is a double well. In the former case we have resonance states, whereas in the latter case we have bound states. 

\section {Quasi-energy spectrum and Floquet bound states}
\begin{figure}[t]
\centering \includegraphics[width=0.8\columnwidth]{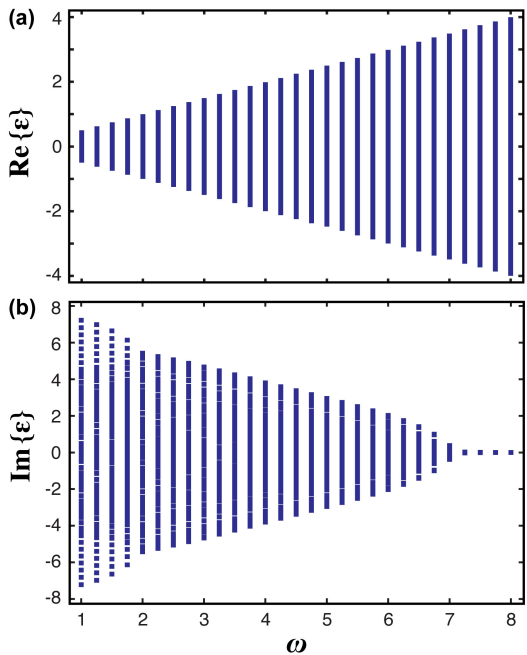}
\caption{(Color online) Real and imaginary parts of the quasi-energies as a function of the modulation frequency $\omega$. The parameters in the potential \eqref{V} are $V_0=9$ and $\beta=0.02$. The real part of $\epsilon$ is taken modulo $\omega$, so that it is always between $-\omega/2$ and $\omega/2$. The number of harmonics is $2N+1=5$.
}\label{epsilon}
\end{figure}

\begin{figure}[tb]
\centering \includegraphics[width=0.9\columnwidth]{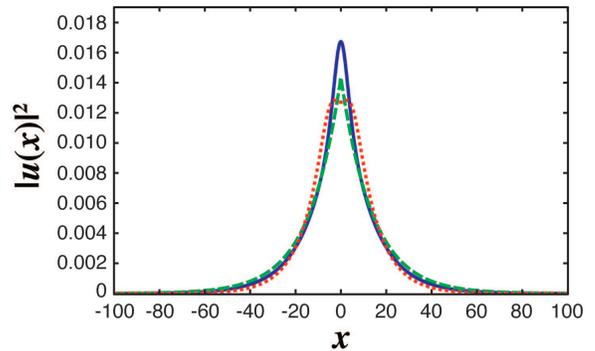}
\caption{(Color online) Probability distribution for the Floquet bound state sustained by $\hat{H}(x,t)$. The three curves correspond to (solid blue curve) the zero-harmonic solution  $|u_0(x)|^2$ for the original time-dependent potential $V(x,t)$, (dotted red curve) the solution for the effective double well potential $V_{\text{eff}}(x)$, and (dashed green curve) the solution for the  $\delta$-function potential well $V_\delta(x)$. The parameter values used in the calculation are $V_0=9$, $\beta=0.02$ and $\omega=10$. The quasi-energy of the bound state is $\epsilon\approx -0.0008$.
}\label{bound}
\end{figure}

From the effective potential description discussed above, we expect the quasi-energy spectrum of the time-periodic Hamiltonian $\hat{H}(x,t)$ to be entirely real-valued at enough high frequencies, despite the Hamiltonian is not Hermitian. On the other hand, at a low modulation frequency the quasi-energy spectrum is expected to be complex valued. Hence 
a transition from a complex to an entirely real quasi-energy spectrum is expected to occur as the modulation frequency $\omega$ is increased.
Since the effective potential description disregards  terms of the order $\mathcal{O}(\omega^{-4})$, it is mandatory to check by a full numerical analysis the transition from a complex to an entirely real quasi-energy spectrum. To numerically compute the quasi-energies $\epsilon$ of $\hat{H}(x,t)$, we look for a solution to Eq.~\eqref{Schr} of the form
\be\label{eq9}
\psi(x,t) = u(x,t)\e^{-i\epsilon t} = \sum_{n=-N}^{N} u_n(x)\e^{i (n\omega -\epsilon) t} ,
\ee
where $n$ runs from the $-N$th harmonic number to the $N$th harmonic number. Substituting this expression into Eq.~\eqref{Schr} yields
\be
(\epsilon - n\omega)u_n(x) = -\half\partial_{x x} u_n(x) + \frac{W(x)}{2}\left[ u_{n-1}(x) + u_{n+1}(x) \right] . 
\ee
The equations above are $2N+1$ coupled time-independent equations, which can be treated as an eigenvalues-eigenvectors problem. They can be solved numerically, for a chosen value of $N$ and after discretization in space, in order to obtain the quasi-energies $\epsilon$ and the harmonic components  $u_n(x)$ of the Floquet eigenstates.
As an example, in Fig.~\ref{epsilon} we show the numerically-computed quasi-energy spectrum of $\hat{H}(x,t)$ versus modulation frequency $\omega$ for a Gaussian potential $W(x)$ [see Eq.~\eqref{V}] for parameter values $V_0=9$ and $\beta=0.02$. The figure clearly shows the transition from a complex to a real quasi-energy spectrum as the modulation frequency is increased above the threshold value $\omega_{th} \simeq 7$. From the computed Floquet eigenstates we also checked the existence of bound states at $\omega> \omega_{th}$. For the parameter values used in the simulations of Fig.~\ref{epsilon}, a single bound state is found, with quasi-energy $\epsilon\approx -0.0008$.
In Fig.~\ref{bound} we show a plot of the probability function of the zeroth-harmonic component of the Floquet bound state for $\omega=10$. For comparison, we computed the energy spectrum of the effective time-independent Hamiltonian $H_{\rm eff}=-(1/2) \partial_{x x}  + V_{\text{eff}}(x)$, and found for the double-well effective potential $V_{\rm eff}(x)$ a single bound state, in agreement with the Floquet analysis. The effective-potential treatment gives a very similar distribution of the probability function of the bound state, as shown in Fig.~\ref{bound}. It should be noted that, despite the effective potential is a double-well potential, it sustains a single bound state, rather than a couple of  bound states as one might expect at first sight. The reason thereof is that, in order to obtain a real energy spectrum, the frequency $\omega$ needs to be large enough. Since $V_{\text{eff}}$ scales as $ \sim 1/ \omega^2$ [see Eq.~\eqref{Vapprox}], the resulting potential well turns out to be very shallow. Hence it can be effectively approximated by a $\delta$-function potential well, i.e.
\be\label{Dirac-delta}
V_{\text{eff}}(x) \approx V_\delta(x) = \alpha\delta(x),
\ee
with $\alpha = \int_{-\infty}^{\infty} V_{\text{eff}}(x) d x$. The $\delta$-function potential well sustains a single bound-state, given by
\be
\psi (x) = 
\left\{ \begin{array}{ll} \sqrt{\mu}\e^{\mu x}, & x<0   \\
\sqrt{\mu}\e^{-\mu x}, & x>0
\end{array} \right. ,
\ee 
where $\mu=-\alpha=\sqrt{-2E}$. In Fig.~\ref{bound} we plot the probability function of this bound state. As one can see, it reproduces very well the spatial profile of the double-well bound state and of the Floquet bound state. 

\section {Optical realization of the imaginary Kapitza pendulum}
The idea of Kapitza stabilization induced by an oscillating purely imaginary potential, and its distinctive feature as compared to an oscillating real potential (the Hermitian case), can find an interesting application and a physical realization in the theory of optical resonators  \cite{Siegman}. In optics and laser physics, non-Hermitian Hamiltonians commonly arise in the description of beam propagation in optical lensguides and resonators with transversely-varying gain and loss media \cite{Si}. Moreover, for short cavities beam propagation in an optical resonator can be mapped into the quantum mechanical Schr\"{o}dinger equation with a potential that is directly related to the mirror profiles \cite{Belanger}.   The basic idea underlying the optical realization of the quantum mechanical Kapitza pendulum is that an optical beam propagating back and forth between two mirrors of an optical resonator mimics the temporal evolution of the wave function of a quantum particle in a potential which is periodically switched between two different values $W_1(x)$ and $W_2(x)$. In this analogy, optical diffraction plays the role of the quantum diffusion (the kinetic energy term in  the Schr\"{o}dinger equation), whereas the transverse profiles and the transversely varying reflectivity of the two mirrors realize the real and imaginary parts, respectively, of the two potentials $W_{1}(x)$ and $W_2(x)$ in the quantum problem.  For $W_2(x)=-W_1(x)=-W(x)$, on average the potential vanishes and basically one retrieves the quantum mechanical formulation of the Kapitza pendulum described in the previous section. 

To formally clarify such an analogy, let us first re-consider the quantum mechanical formulation of the Kapitza pendulum  with a square-wave (rather than sinusoidal) modulation in time of the potential, i.e. let us assume
\begin{equation}
V(x,t)=W(x)F(\omega t)
\end{equation}  
where $F(\xi)$ is the square-wave function with period $2 \pi$ [$F(\xi+2 \pi)=F(\xi)$, $F(\xi)=1$ for $0<\xi< \pi$ and $F(\xi)=-1$ for $\pi<\xi<2 \pi$], and $\omega$ is the modulation frequency. Indicating by $\psi(x,t)=u(x,t) \exp(-i \epsilon t)$ the Floquet eigenstate of the Schr\"{o}dinger equation with quasi energy $\epsilon$ [see Eq.~\eqref{eq9}], it can be readily shown that for a square-wave modulation in time the function $u(x,0) \equiv u(x)$ satisfies the following equation
\begin{equation}\label{eq14}
\exp(-i \hat{H}_2 T/2) \exp(-i \hat{H}_1 T/2) u(x) = \exp(-i \epsilon T) u(x)
\end{equation}
where $T=2 \pi / \omega$ is the modulation period, $\hat{H}_1=-(1/2) \partial_{xx}+W(x)$ and $\hat{H}_2=-(1/2) \partial_{xx}-W(x)$. The eigenvalue equation \eqref{eq14} is exact and holds for any value of the modulation frequency. In the large modulation frequency limit, i.e. for $T \rightarrow 0$, Eq.~\eqref{eq14} greatly simplifies by application of the Baker-Campbell-Hausdorff formula or, equivalently, by Taylor expansion of the exponential operators. The analysis requires to push the asymptotic expansion up to the order $\sim T^3$. Using the identity 
\begin{widetext}
\begin{equation}
\exp(\hat{X}) \exp(\hat{Y}) \simeq  \exp \left\{ \hat{X}+ \hat{Y}+\frac{1}{2} [\hat{X},\hat{Y}]+ \frac{1}{12} \left[ \hat{X}-\hat{Y},[\hat{X},\hat{Y}] \right] \right\}.
\end{equation}
\end{widetext}
with $\hat{X}=-i\hat{H_2}T/2$ and $\hat{Y}=-i\hat{H_2}T/2$, it then follows that the eigenvalue equation \eqref{eq14} is satisfied for 
\begin{eqnarray}\label{eq16}
\epsilon u(x) & = & \frac{1}{2}(\hat{H}_1+\hat{H}_2) u(x)+i \frac{T}{8} [ \hat{H}_1, \hat{H}_2] u(x) \nonumber \\
& - & \frac{T^2}{96} \left[ \hat{H}_1-\hat{H}_2, [ \hat{H}_1,\hat{H}_2]   \right] u(x).
\end{eqnarray}
After computation of the commutators entering on the right hand side of Eq.~\eqref{eq16}, after some cumbersome calculations one obtains
 \begin{equation}\label{eq17}
 -\frac{1}{2} \partial_{xx}u+i\frac{T}{8} \left( \partial_{xx}W u+2 \partial_x W \partial_x u \right)+\frac{T^2}{24} (\partial_x W)^2 u= \epsilon u.
 \end{equation}
The effective potential formulation in the high-frequency regime, discussed in the previous sections, is obtained from Eq.~\eqref{eq17} after introduction of the new function $y(x)=u(x) \exp\{ -i(T/4) W(x) \} $, which satisfies the stationary Schr\"{o}dinger equation
\begin{equation}\label{eq18}
-\frac{1}{2} \partial_{xx}y+V_{\rm{eff}}(x)y= \epsilon y
\end{equation}
 with an effective potential given by 
 \begin{equation}\label{eq19}
 V_{\rm{eff}}(x)=\frac{T^2}{96} \left( \frac{\partial W}{\partial x} \right)^2=\frac{4 \pi^2}{96 \omega^2} \left( \frac{\partial W}{\partial x} \right)^2.
 \end{equation}
  Note that the form of the effective potential given by Eq.~\eqref{eq19} differs from Eq.~\eqref{Vapprox} just for a multiplication factor. The reason thereof is the different temporal modulation of the potential, square wave versus sinusoidal.\par Let us now outline the optical resonator analogue of the quantum mechanical problem described above. We consider paraxial light beam propagation at wavelength $\lambda$ back and forth between two mirrors of a Fabry-Perot cavity. Like in Ref.~\cite{Belanger}, we assume a one transverse dimension, i.e. a slab geometry. In each cavity round trip, the optical field amplitude $u(x)$ at a reference plane (for example at the right mirror plane) changes because of diffraction in the propagative region between the two mirrors, and because of the reflection from the mirrors.  Mirror reflection can introduce a transversely-varying phase (for non-flat mirrors) and/or a transversely-varying intensity reflection (for variable-reflectivity mirrors), see e.g. \cite{Siegman}. Let us first consider the former case, i.e. perfectly reflecting mirrors with non-flat and generally aspherical surfaces, see Fig.~\ref{fig5}(a). As it will be shown below, this system realizes the ordinary Kapitza pendulum with a real potential. A resonator mode, by definition, is a field distribution $u(x)$ that reproduces itself after one cavity round trip, apart from a multiplication factor. The resonator mode profiles $u(x)$ are thus found as solutions of the eigenvalue equation (see, for instance \cite{Belanger})
 \begin{eqnarray}\label{eq20}
 \left\{ \exp (\mathcal{D}) \exp(iW_1(x)) \exp(\mathcal{D}) \exp(iW_2(x))  \right\} u(x)  & = &  \nonumber \\
\exp(-i \mu) u(x) &  &
 \end{eqnarray} 
 with eigenvalue $\exp(-i \mu)$. In Eq.~\eqref{eq20}, $\mathcal{D}=i (d/2k) \partial^2_{xx}$ is the one-way diffraction operator, $d$ is the mirror spacing, $k= 2 \pi / \lambda$ is the wave number of light and $W_{1,2}(x)$ are the phase delays introduced by the curved mirrors. The latter are simply related to the geometric profiles of the mirror surfaces \cite{Belanger} via the relation $W_{1,2}(x)=k\Delta_{1,2}(x)$, were $\Delta_{1,2}(x)$ is the distance (with sign) of the curved surface of the mirror from the reference flat surface [see Fig.~\ref{fig5}(a)]. In particular, for $\Delta_{2}(x)=-\Delta_1(x)$, one has $W_2(x)=-W_1(x) \equiv -W(x)$ and Eq.~\eqref{eq20} reads
 \begin{eqnarray}\label{eq21}
 \left\{ \exp (\mathcal{D}) \exp(iW(x)) \exp(\mathcal{D}) \exp(-iW(x))  \right\} u(x)  & = &  \nonumber \\
\exp(-i \mu) u(x) &  &
 \end{eqnarray} 
 In this case,  assuming the short cavity limit $d \rightarrow 0$ the mode $u(x)$ undergoes a slight change over one cavity round trip \cite{Belanger}, and one may expand the operator $\exp(\mathcal{D})$ 
 in Taylor series up to first order, i.e one may set   $\exp(\mathcal{D}) \simeq 1+\mathcal{D}=1+id/(2 k) \partial_{xx}$ in Eq.~\eqref{eq21}. Similarly, the eigenvalue is close to one, and thus one can set $\exp(-i \mu) \simeq 1-i \mu$. Under such approximations, Eq.~\eqref{eq21} reads
 \begin{equation}\label{eq22}
 -\frac{d}{k} \partial_{xx} u+\frac{d}{2k} \left\{ 2 i \partial_x W \partial_x +i \partial_{xx}W +(\partial_x W)^2 \right\} u = \mu u.
 \end{equation}
  The eigenvalue equation \eqref{eq22} can be cast into the standard Schr\"{o}dinger form after the change of function $u(x)=y(x) \exp[i W(x)/2]$. This transforms Eq.~\eqref{eq22} into  the Schr\"{o}dinger equation \eqref{eq18} for the function $y(x)$, with $\epsilon=k \mu/(2d)$ and with the effective potential
  \begin{equation}
  V_{\rm {eff}}=\frac{1}{8} \left( \frac{\partial W}{\partial x}\right)^2.
  \end{equation}
\begin{figure}[t]
\centering \includegraphics[width=0.8\columnwidth]{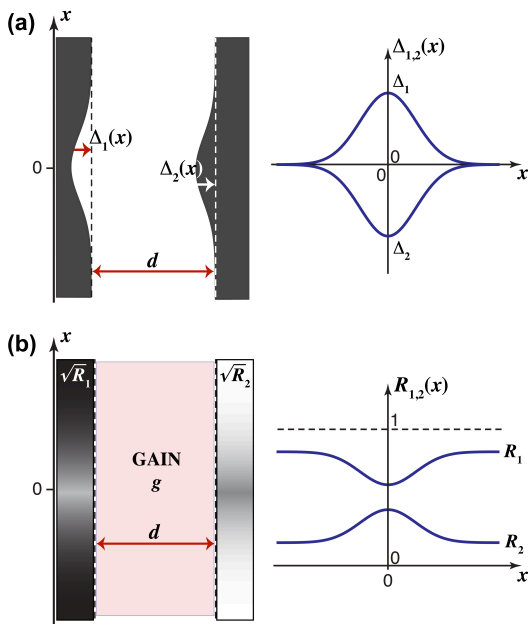}
\caption{(Color online) Optical resonator implementation of the quantum mechanical Kapitza pendulum for (a) a real, and (b) an imaginary potential $W(x)$. In (a) the Fabry-Perot resonator consists of two perfectly reflecting mirrors  with curved surfaces and with $|\Delta_1(x)|=|\Delta_2(x)|$. In (b) the resonator is made by two flat mirrors with transversely-varying reflectances $R_{1,2}(x)$, filled by a gain medium. A typical behavior of $R_{1,2}(x)$ is shown in the inset. Mirror spacing is $d$.}\label{fig5}
\end{figure}
  Hence a beam bouncing back and forth between two fully reflecting and aspherical mirrors, which introduce opposite delays $W(x)$ and $-W(x)$ of the wave front, basically mimics the quantum Kapitza pendulum with a real potential periodically switching between the two values $W(x)$ and $-W(x)$.\\
  To realize the resonator analogue of the imaginary Kapitza pendulum, let us consider the resonator of Fig.~\ref{fig5}(b), which is composed by two flat mirrors with transversely-varying reflectivities $\sqrt{R_1(x)}$ and $\sqrt{R_2(x)}$, where $0<R_{1,2}(x) \leq 1$ are the mirror reflectances \cite{noteuffa}. Since the resonator is lossy, to sustain a stationary mode we include a gain medium that fills the resonator. The gain medium provides a uniform (saturated) gain per unit length equal to $g/2$. As compared to the case of Fig.~\ref{fig5}(a), a beam bouncing back and forth between the two mirrors experiences, in addition to diffraction in the propagative region, uniform amplification in the gain medium and transversely-varying losses at the mirrors. The resonator mode profiles $u(x)$ now satisfy the following eigenvalue equation [compare with Eq.~\eqref{eq20}]
  \begin{eqnarray}\label{eq24}
 \left\{ \exp (\mathcal{D}) \sqrt{R_1(x)}  \exp(\mathcal{D}) \sqrt{R_2(x)}  \exp(gd)  \right\}    u(x) &  &  \nonumber \\
 = \exp(-i \mu) u(x). \;\;\;\; &  &
 \end{eqnarray} 
  Let us now assume that the reflectance profiles $R_1(x)$ and $R_2(x)$ satisfy the following constraint
  \begin{equation}\label{eq25}
  R_1(x) R_2(x)=\exp(-2gd)
  \end{equation}
  with $R_{1,2}(x)$ taking constant values $R_{1,2}^{\infty}$ at $x \rightarrow \pm \infty$ [see the inset of Fig.~\ref{fig5}(b)]. 
  This basically means that, if for instance $R_1(x)$ has a well around $x=0$, than $R_2(x)$ has a  hump around $x=0$; see Fig.~\ref{fig5}(b). Under such a condition, the eigenvalue equation \eqref{eq24} takes the form given by Eq.~\eqref{eq21} with the replacement 
  \begin{equation}
  W(x) \rightarrow -i \; {\rm ln} \sqrt{R_1(x)}
  \end{equation}
This means that the transverse modes of the resonator of Fig.~\ref{fig5}(b) with variable reflectivity mirrors satisfying the condition \eqref{eq25} are found by solving the Schr\"{o}dinger equation \eqref{eq18} with the effective potential
\begin{equation}
V_{\rm {eff}}=-\frac{1}{8} \left( \frac{\partial {\rm ln} \sqrt{ R_1(x)}} {\partial x}\right)^2
\end{equation}
which realizes the quantum imaginary Kapitza pendulum, owing to the reversal of the sign in the effective potantial.\par
The different stabilization properties of the real and imaginary Kapitza pendulum, discussed in the previous sections, have a strong impact into the stability of resonators that trap light using either variable phase [Fig.~\ref{fig5}(a)] or amplitude [Fig.~\ref{fig5}(b)] mirrors.  In the former case, for aspherical mirrors which are asymptotically flat (i.e. $\Delta(x) \rightarrow 0$ as $x \rightarrow \pm \infty$) the resonator turns out to be always unstable, i.e. it does not sustain truly stationary and confined  electromagnetic modes, but only leaky modes owing to the shape of the effective potential. Conversely, in the latter case the resonator with variable-reflectivity mirrors can sustain stationary and confined  electromagnetic modes, i.e. it is stable.

\section {Conclusions}
In this work we have proposed a generalization of the Kapitza stabilization effect to imaginary potentials. In the classical case, it has been shown that an imaginary oscillating inertial force introduces two stable fixed points, which deviate from the usual vertical positions of the pendulum. 
In the quantum case we have shown that a frequency-modulated purely imaginary potential may lead to an entire real quasi-energy spectrum of the non-Hermitian Hamiltonian for a large modulation frequency, with a transition from a complex to a real spectrum as the modulation frequency is increased. In such a regime we found that Floquet bound states can be sustained by an oscillating imaginary potential. This is a very distinctive feature as compared to the quantum Kapitza stabilization in the Hermitian case, where stabilization is imperfect and resonance states (rather than truly bound states) can be sustained. An application of the imaginary Kapitza pendulum to the stability properties of optical resonators with variable-reflectivity mirrors has been suggested. Our results indicate that  stabilization in classical and quantum systems by oscillating potentials show a very distinctive behavior when the potential is allowed to become imaginary, and motivate further studies on the general properties of driven non-Hermitian systems. For example,  Kapitza stabilization for imaginary potentials could be of relevance to optics in media with gain and loss regions \cite{OP}, where Kapitza stabilization might provide an unexpected mechanism of light guiding and trapping  \cite{Ciattoni}, and to open quantum systems, such as open two-level atomic systems interacting with near resonant light, where the dynamics can be described by an effective linear Schr\"{o}dinger equation with a complex potential \cite{MW}. Our study might be also of interest to the broad field of quantum simulations, where  realistic proposals to implement non-Hamiltonian (either non-dissipative or dissipative)  wave equations with atoms, ions, molecules and superconducting quantum circuits are currently under active  investigation \cite{NP10,review,Casanova,Tureci}.


\section{Acknowledgement}
This work was supported by the Fondazione Cariplo (Grant No. 2011-0338).



\begin{thebibliography}{99}
\bibitem{note} We exclude the unlikely possibility of bound states embedded in the continuum of scattered stases, which might occur for very special profiles of $W(x)$.
\bibitem{noteuffa}
 We note that  mirrors with radially-varying reflectance are routinely used in laser design with unstable resonators.
See, for instance, Sec. 23.3 of Ref.~\cite{Siegman} and references therein.
\end{thebibliography}
\end{document}